# What limits performance of weakly supervised deep learning for chest CT classification?

**Fakrul Islam Tushar[a,b], Vincent M. D'Anniballe[a], Geoffrey D. Rubin[c], Joseph Y. Lo[a,b]**
[a]*Center for Virtual Imaging Trials, Carl E. Ravin Advanced Imaging Laboratories, Department of Radiology, Duke University School of Medicine, Durham, NC.*
[b]*Department of Electrical and Computer Engineering, Duke University, Durham, NC.*
[c]*Department of Medical Imaging, University of Arizona, Tucson AZ.*

**Abstract:**

Weakly supervised learning with noisy data has drawn attention in the medical imaging community due to the sparsity of high-quality disease labels. However, little is known about the limitations of such weakly supervised learning and the effect of these constraints on disease classification performance.

In this paper, we test the effects of such weak supervision by examining model tolerance for three conditions. First, we examined model tolerance for noisy data by incrementally increasing error in the labels within the training data. Second, we assessed the impact of dataset size by varying the amount of training data. Third, we compared performance differences between binary and multi-label classification.

Results demonstrated that the model could endure up to 10% added label error before experiencing a decline in disease classification performance. Disease classification performance steadily rose as the amount of training data was increased for all disease classes, before experiencing a plateau in performance at 75% of training data. Last, the binary model outperformed the multilabel model in every disease category. However, such interpretations may be misleading, as the binary model was heavily influenced by co-occurring diseases and may not have learned the specific features of the disease in the image.

In conclusion, this study may help the medical imaging community understand the benefits and risks of weak supervision with noisy labels. Such studies demonstrate the need to build diverse, large-scale datasets and to develop explainable and responsible AI.

## 1. Introduction:

Medical imaging has experienced a widespread application of deep convolutional models for various disease classification and segmentation tasks [1-8]. Despite high performance, these models are narrow in scope, with the majority targeting one disease or organ, which does not address the co-occurrence of multiple diseases often found in the same image study. This is primarily due to the prohibitive expense and inherent variability for high throughput annotation of multiple organs and diseases.

In response to these limitations, recent computer-aided diagnosis models have leveraged weak, noisy text annotations extracted from radiology reports [9-15]. The approach used to mine the unstructured text for relevant disease label information relies on rule-based algorithms (RBA) or natural language processing (NLP). RBAs implement logical, conditional operations to predict labels from reports and can be adjusted to tackle new tasks. However, the sheer diversity within radiology text reports (e.g., writing style, grammatical errors, negations) limits the performance and generalizability of strict coding rules [16, 17]. However, different from RBAs, NLP based models have demonstrated superior potential to be applicable in various radiology text analysis tasks [18-23].

Using automatic label extraction guided by an RBA, Aurelia *et al.* reported label accuracy of 0.93 Micro-F1 score using a recurrent neural network in the Padchest dataset [15]. Wang *et al.* used the ChestX-ray8 dataset to automatically extract labels using natural language processing with labeling accuracy > 90% [9]. Draelos et al. reported an average of 0.97 F-score labeling radiology reports using an RBA encompassing over 80 diseases [13]. Finally, Tushar et al. extracted labels for multiple diseases of different organ systems (lungs/pleura, liver/gallbladder, and kidneys/ureters) and reported RBA labels were 91 to 99% accurate [14]. In a follow-up study D'Anniballe *et al.* trained an RNN model utilizing the RBA labels, and final RBA were equal to or greater than both previously reported RBAs for all diseases, with accuracy ranging from 91% to 99% and F-score from 0.85 to 0.98 [24]. In spite of reported high labeling accuracy, most studies acknowledge label noise as a limitation. There is a need for a rigorous investigation into the effect of such labeling noise on classification performance.

The amount of data used to train is another crucial aspect of deep learning. Sun et al. demonstrated that vision task performance increased logarithmically based on training data size [25]. In medical imaging, Yan et al. reported that disease classification performance improved steadily as additional training cases were used from the DeepLesion dataset [12]. These studies require large datasets at the scale of health systems and are thus very rare.

Lastly, image classifiers can produce several types of outcomes: binary, multi-label, or multi-class. Binary models classify images as either positive or negative for disease. Multi-label models instead determine occurrence of several diseases simultaneously, while multi-class models determine one of multiple diseases that are exclusive from others. Co-occurrence of disease is a natural scenario in medical imaging data. The possibility to classify multiple disease labels simultaneously more closely mirrors the task of a radiologist compared to a binary classification system. However, this introduces new challenges to understand if a model is learning the correct disease features rather than being influenced by co-occurring disease features.

This work explored these unanswered questions of weak supervision for multi-label medical data using only case-level labels extracted from radiology reports. We focused this study on chest CT as a common medical imaging study. The main contributions of this paper are threefold:

1. Using labels from three generations of RBA algorithms from the same group [24, 26, 27], we have isolated how different levels of label noise affect deep learning models' performance. We have further incrementally introduced random label noise to multiple and single co-occurrence classes to analyze the effects.
2. We have leveraged the large amount of data to examine the impact of training data size on model performance.
3. We have investigated differences between binary and multi-label classification on a multi-label dataset by measuring the influence of disease co-occurrence on binary classification.

## 2. Materials and Methods:

### 2. 1. Dataset and Preprocessing:

Image data and labels were derived from previous studies with IRB approval. An RBA was used to extract 7,441 case-level disease labels of the lungs/pleura from the text reports of 5,044 body CT scans of 4,639 unique patients from Duke Health [14, 24]. The rules of the RBA were built upon general relationships of anatomy and disease keywords. For example, if a sentence contained the organ descriptor "lung" or "lobe" and abnormality "nodule" and there was no negative like "no" or "without," then the report was positive for lung nodule.

The labels are considered weak because they relate to the entire CT volume without location information and are noisy because the rule-based decisions are imperfect. Manual validation of 1154 rule-based labels confirmed they were at least 92% accurate [14, 24]. The lungs/pleura were labeled as no apparent disease versus having one or more of four diseases: atelectasis, nodule, emphysema, and/or effusion. Details of the RBA including the complete keyword dictionary structure were previously described together with a public code repository [24].

The total number of volumes associated with each disease and the no apparent disease label used in classification tasks was 1194 for emphysema, 1465 for effusion, 1628 for nodule, 1758 for atelectasis, and 1396 for no apparent disease. Fig. **1** illustrates the co-occurrence and association between diseases for this multi-label data set.

Prior to experimentation, all CT volumes were resampled to voxels of size 2 *mm* x 2 *mm* x 2 *mm* via B-spline interpolations, clipped to intensity range (-1000, 800) HU, and normalized to 0 mean and 1 standard deviation. The CT volumes were randomly divided into subsets to train (70%), validate (15%), and test (15%) the image classification model. Splitting was performed by subject and separately for no apparent disease vs. diseased classes to preserve disease prevalence across each subset.

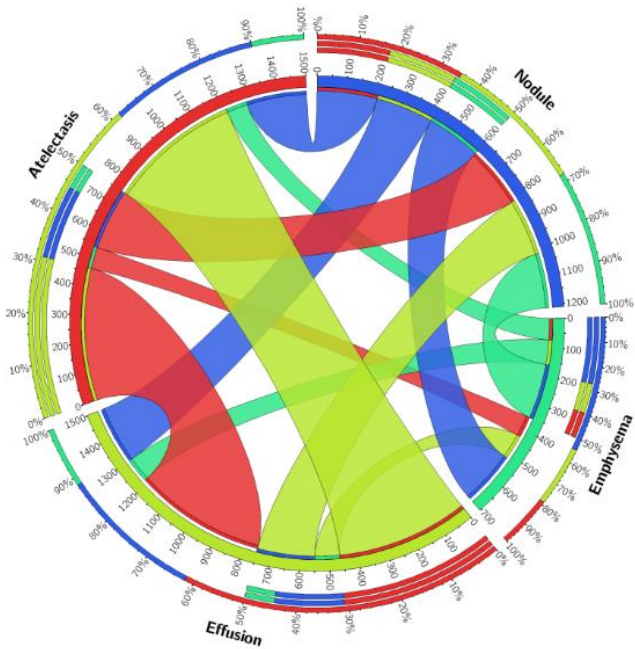

**Figure 1.** Chord diagram displays the occurrence and co-occurrence of targeted diseases in the lungs/pleura among unique subjects.

## 2.2. Classification Workflow:

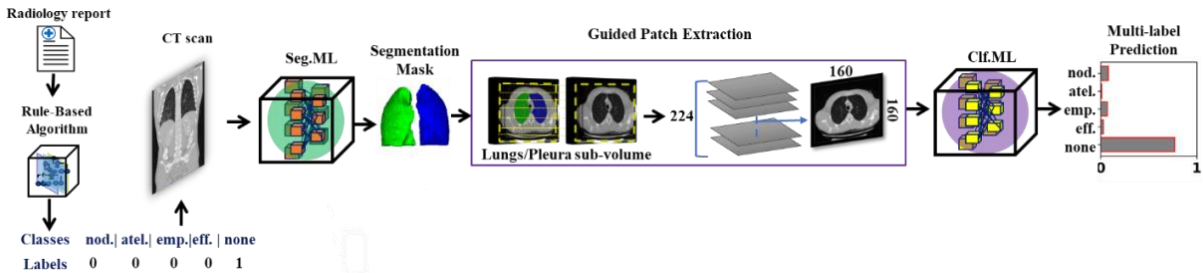

**Figure 2.** Study overview. Illustration of weakly-**supervised** classification study workflow. 4 abnormalities and no apparent disease labels were extracted using RBAs [24, 26, 27], a segmentation model was used to extracted lung patches from body CTs and finally we trained and validated 3D CNN for classification on 3D CT volumes to predict all 4 abnormalities and no apparent disease simultaneously **or in binary setup (diseased vs. no apparent of disease)**.

Fig. 2 displays the overall classification workflow. First, applying an RBA to each radiology report associated with a CT scan allowed disease label acquisition [24, 27, 28]. Afterward a segmentation module was trained in semi-supervised manner, which provided a lungs segmentation mask for guided patch extraction for classification. DenseVNet was used as segmentation model and in first stage trained with CT volumes with organ margins labeled on [29]. each transverse section originally developed for the 4D extended cardiac-torso (XCAT) phantom [30] and afterward was fine-tuned using 30 additional randomly selected, diseased lung cases [14].

The 3D CNN used for multi-label classification in this study was introduced in our prior study [14]. Classification CNN is a 4 resolution 3D CNN inspired by resnet [14, 28]. One initial convolution was performed on input volumes, afterward, features were learned in three resolution scales using three R-block units in each resolution. A R-block consists of batch-normalization, rectified linear unit (ReLu) activation, and 3D convolution [14]. We used 3x3 kernels that are faster to convolve with and contain fewer weights. Batch-normalization allows normalization of the feature map activation at every optimization step. After each resolution, the features were reduced half and the number of filters (16, 32, 64, 128) is doubled. After the 3rd resolution last R-block features passed through batch-normalization, ReLu followed by a global average-pooling and Dropout (0.5), finally sigmoid classification layer for the final multi-label prediction or binary prediction. Weighted cross-entropy loss (W-CEL) was used as loss function [9]. All models were trained from scratch using Adam optimizer with a learning rate of 0.0001 and classification performance was based on the receiver operating characteristics (ROC) area under the curve (AUC).

### 2.3. Label Noise Experiments:

The RBAs used in our study underwent rigorous development phases initiated by Han *et al.*, further refined by Faryna *et al.*, and finally Tushar *et al.* refined the RBA further [14, 26, 27].

D'Anniballe *et al.* in explained details of the RBA, compared the above RBAs, and showed a significant difference in label accuracy by RBAs against a test set [24].

In this study we have explored all three RBAs, each with appreciable improvement in accuracy [24, 26, 27]. To examine the effect of label noise on model performance, each phase was used to extract labels from the training data and then subject to comparison of the manually labeled test set. All hyperparameters otherwise remained the same.

Based on the performance of different development RBAs we have observed different level of label noise to different classes [24, 26, 27]. This different level of noise in different classes may not show the co-occurrence effect to each class in multi-label as the noise level is different for each class. To address this issue and analyze each class's performance nature with equal amount of noise in labels, we randomly introduced different percentage of noise, e.g., 10%, 20%, 25% and 50%, to labels from the RBA by D'Anniballe *et al.* to training data and trained different combinations, such as changing all 4 four disease class labels to changing labels for each disease class separately [24]. Firstly, we chose lowest performing class nodule and highest performing class and later we changed one class at a time to observe that label noise effect to any other co-occurrence diseases.

**2.4. Analysis of the model performance with different amounts of data:**

To analyze the effect of the model performance associated with dataset size, different numbers of samples were used for training the multi-label classifier. The training dataset was randomly reduced to 5%, 10%, 25%, 50%, and 75% to train the model keeping all other hyperparameters for training.

**2.5. Binary and Multi-label experiments:**

Our dataset suffers from high co-occurrence and association between diseases shown in Fig. 1, which is a natural scenario in the medical scenario. With our dataset, we have compared multi-label classification (MLCL) and binary classification (BCL) and performed a series of experiments to understand and analyze the effect of co-occurrence in performance. Fig. 3 shown the classification experiment performed.

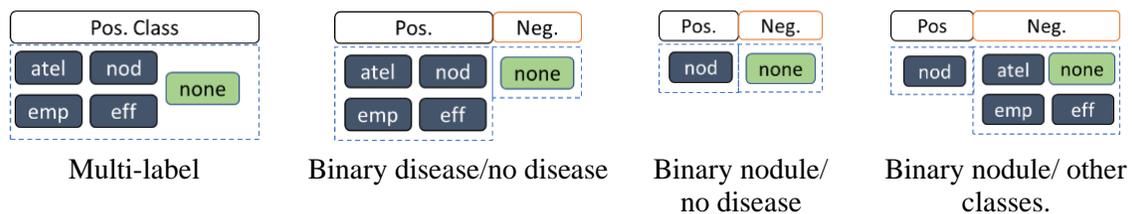

Multi-label     Binary disease/no disease     Binary nodule/ no disease     Binary nodule/ other classes.

**Figure 3**. Four classification models were created based on which classes are positive vs. negative to analyze the effect of co-occurrence on performance. Multi-label: Simultaneously classifying all four disease classes and no apparent disease class; Binary disease / no disease: no disease vs. combination of all 4 diseases; Binary nodule / no disease: nodule vs. no disease; and Binary nodule / other classes: nodule vs. other classes. Nodule, atelectasis, emphysema, effusion, and no apparent disease were denoted as nod., atel., emp., eff., and none.

In the **multi-label classification** approach, the CNN model needs to simultaneously predict all four disease classes: atelectasis, nodule, effusion, emphysema, and no apparent disease class. To further understand the multi-label co-occurrence influence, we have computed the AUC of each disease classed based on co-occurrence. In **binary disease/no disease,** the model was trained using no apparent disease vs. abnormalities combining disease classes atelectasis, nodule, emphysema, and effusion to an abnormal class against no apparent disease class.

To analyze the hardship of the model to distinguish nodule features from all other non-nodule features, we have designed the experiment **binary nodule/other classes.** In this experiment, we classified nodule (All nodule positive) and other classes (combined atelectasis, emphysema, effusion, and no apparent disease). By nature, this and multi-label setup have the same difficulties learning and differentiating nodule vs. non-nodule features.

## 3. Results:

The test set consists of 771 CT volumes from 771 unique subjects with 1154 labels for lungs/pleura. All 1,154 labels in the test set were manually validated. The number of positive examples for each class were as follows: 251 for atelectasis, 296 for nodule, 193 for emphysema, 205 for effusion, and 209 for no apparent disease. The RBA by D'Anniballe *et al.* [24] performed highest with accuracy from 92% to 99%, and F-score from 0.89 to 0.98, whereas RBA by Han *et al.* [26] performed lowest with accuracy from 0.77% to 0.86%, and F-score from 0.45 to 0.74, shown in fig. 4(a).

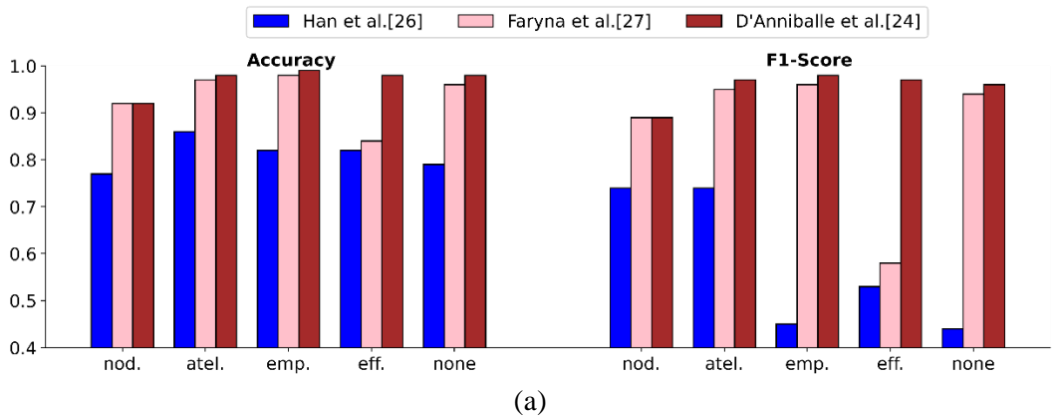

(a)

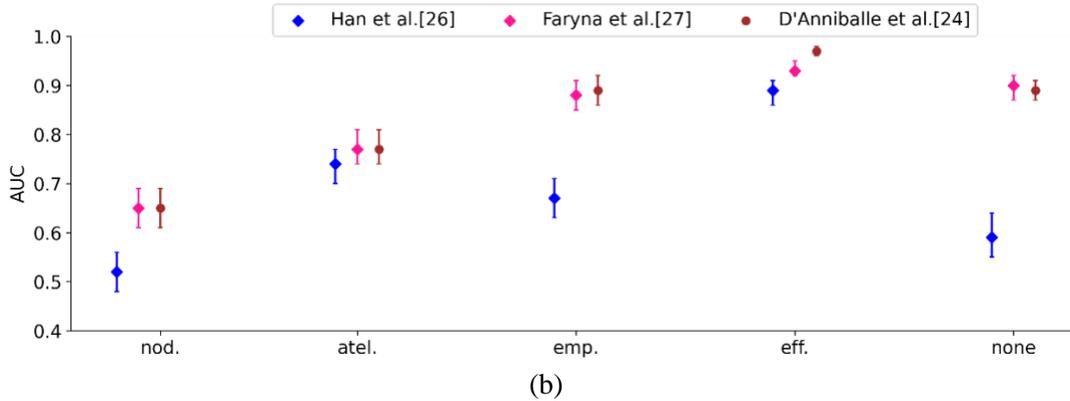

(b)

**Figure 4.** (a) RBAs performance on the test set (b) Performance of the multi-label classification models for lung/pleura with different levels of label noise introduced by RBA of different development phases. Error Bar represents the 95% confidence interval (CI). RBA by Faryna *et al.[27] and* D'Anniballe *et al.[24]* performed similarly and outperformed RBA by Han et al.[26]. Using the Labels from RBAs while training image classifier showed similar trends as RBA performance. Nodule, atelectasis, emphysema, effusion, and no apparent disease were denoted as nod., atel., emp., eff.,and none. respectively.

Fig. 4(b) shown the performance of the multi-label CT classification models for lung/pleura with different levels of label noise introduced by RBAs [24, 26, 27] of different development phases. A significant performance improvement has been observed in a multi-label model trained with RBA by D'Anniballe *et al.* to Han *et al.* [24, 26]. RBA labels by D'Anniballe *et al.* trained multi-label classification model performed AUC was > 0.85 for effusion and emphysema, moderate for atelectasis with AUC 0.77, but poor or nodules with AUC of 0.65 [24]. RBA labels by Faryna *et al.* and RBA labels by D'Anniballe *et al.* trained classification model showed similar performance except for effusion [24, 27]. Whereas the multi-label model trained with labels from RBA by Han *et al.* performed comparatively worst with classification AUC> 0.60 for nodule and no apparent disease, moderate for atelectasis and emphysema with no apparent disease AUC > 0.67, and respectable for effusion with 0.89 [26].

Fig. 5 (a) to (f) shown the performance of the multi-label classification for lung/pleura trained with different level of random noise added to labels extracted using RBA from D'Anniballe *et al.[24]*. The results showed a similar trend as up to 10% added noise to multiple labels and up to 20% added noise to a single label.

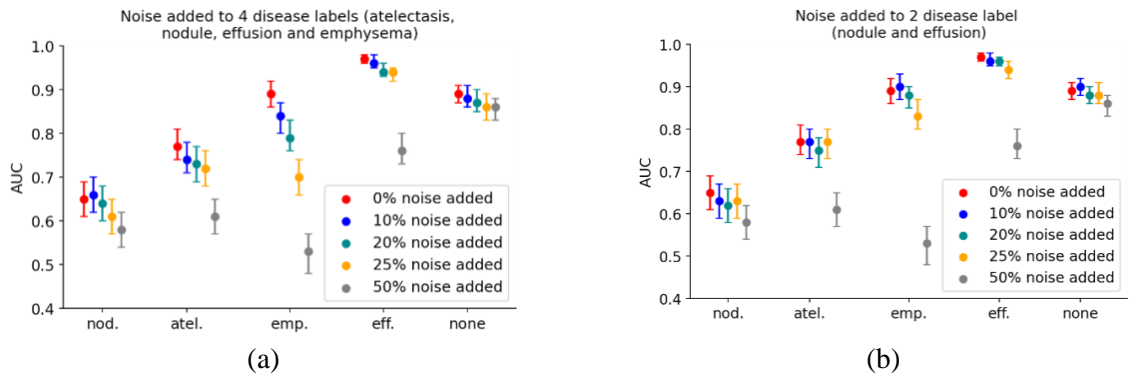

(a)　　　　　　　　　　　　　　　　　　　(b)

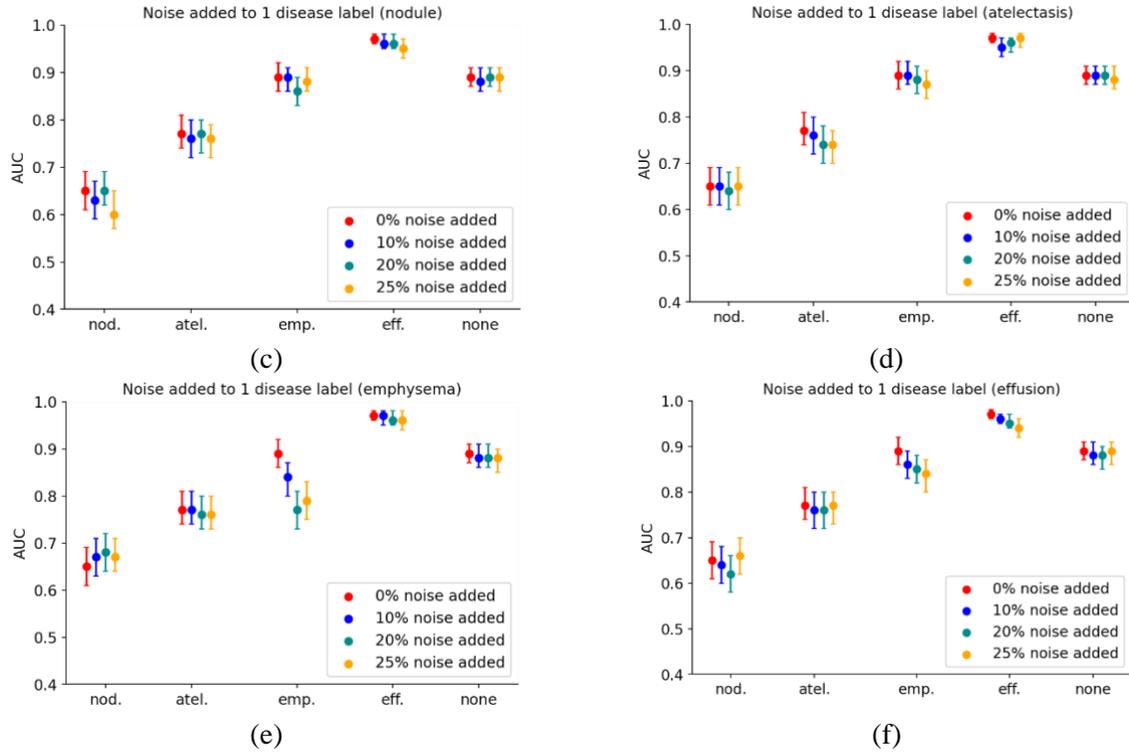

**Figure 5.** Classification performance on test-set with different proportions of label error added to training data. Error Bar represents the 95% confidence interval (CI). Nodule, atelectasis, emphysema, effusion, and no apparent disease were denoted as nod., atel., emp., eff. and none. respectively.

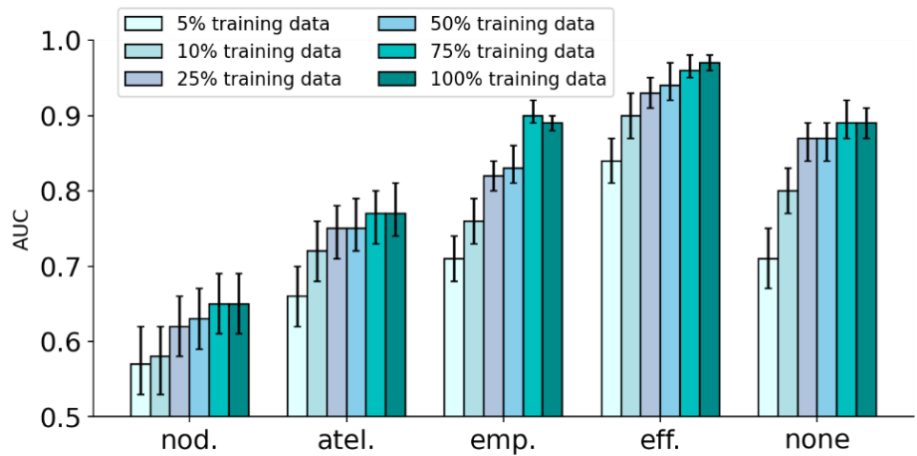

**Figure 6.** AUC of classification performance on test-set when different proportions of training data are used. Error Bar represents the 95% confidence interval (CI). Nodule, atelectasis, emphysema, effusion, and no apparent disease were denoted as nod., atel., emp., eff. and none. respectively.

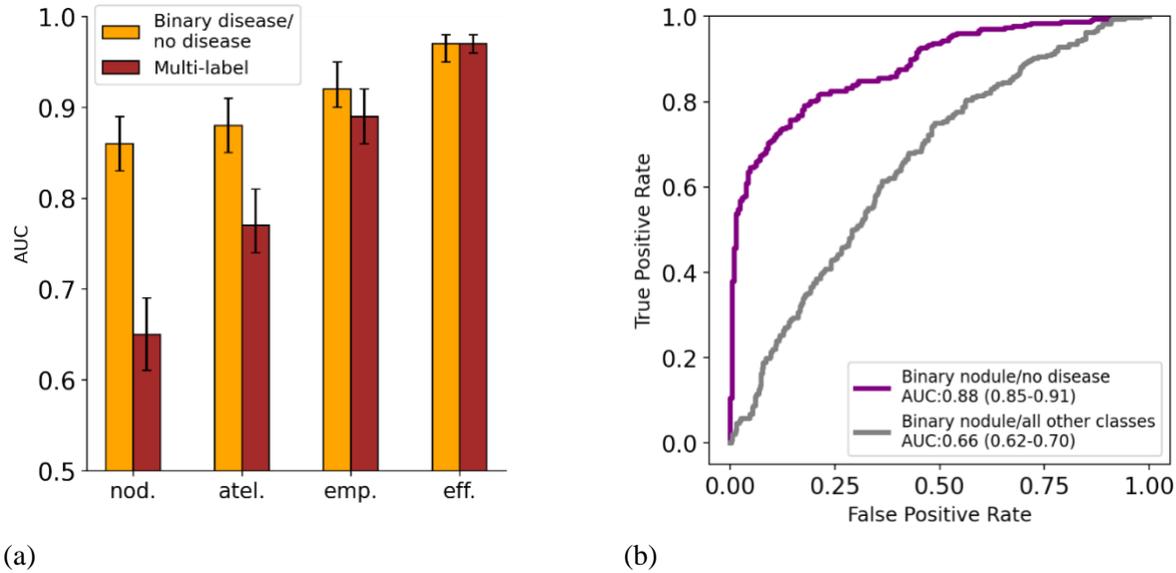

(a) (b)

**Figure 7. (a)** Classification performance on test-set of binary and multi-label classification. Error Bar represents the 95% confidence interval (CI). **(b)** Classification performance of Nodule Class with different experiments. Nodule, atelectasis, emphysema, effusion and no apparent disease were denoted as nod., atel., emp., eff., and none. respectively.

Fig. 6 shown the steady performance improved for all the cases as added more training data. Fig. 7(a) shown the performance of the binary disease/no disease and multi-label classifications. For Emphysema, Effusion binary and multi-label model performed similarly. In contrast, there is a significant performance increase in the case of binary disease/no disease classification compared to multi-label for nodule and atelectasis.

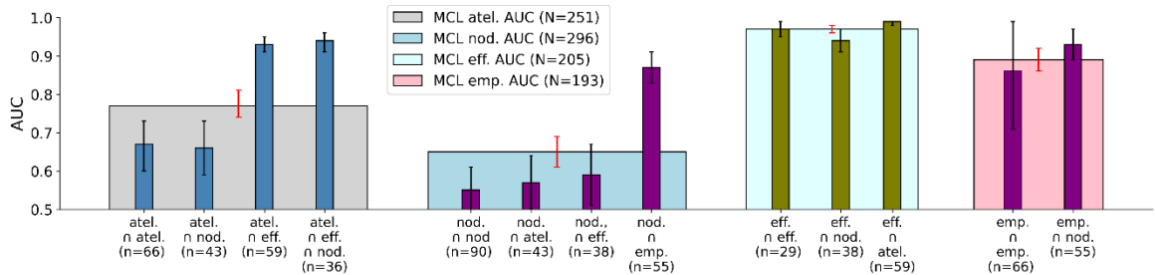

**Figure. 8:** Performance of the multi-label classification (MCL) model on the test set based on the single, two and three class major associations. **N**=number of total positive volumes associated to that class, **n**=numbers of positive volume associated with the association; error bar represents 95% confidence interval. Nodule, atelectasis, emphysema, and effusion were denoted as nod., atel., emp., and eff. . respectively.

Fig 8. Shown the performance of the disease classes in terms of exclusive and co-occurrence with other classes. Nodule class co-occurrence with emphysema always increases performance > 0.80, wherewith other classes its performance poorly < 0.60. Similar relation can be observed between atelectasis and effusion, where atelectasis performance is higher when co-occurred with effusion.

## 4. Discussion:

In this work, we explored the various aspects of weakly supervised classification using multi-label chest CT data by exploring error tolerance, examining the impact of dataset size, and investigating differences between binary-and multi-label classifiers.

Noisy labels are unavoidable in the case of automated label extraction, especially when using complex radiology reports. Researchers often report this inherent noise as a limitation but do not examine the direct implications of this error. In the present study, we attempted to understand effect of such error in model performance and learning. Specifically, we examined the effect of label noise using different RBAs by Han *et al.*[26], Faryna et al.[27] and D'Anniballe *et al.* [24]. Additionally, we have randomly added label noise to multiple and single disease classes.

We observed a significant improvement in image classification performance using the labels from our RBA by D'Anniballe *et al.*[24] (which contained added rules for length, wild card organ descriptors, and handling of typos) compared to RBA by Han *et al. [26]* . Furthermore, these error tolerance experiments suggested the effect of label noise is minimum when noise is introduced to a single disease class compared to n multiple classes. Moreover, we demonstrated that each disease class reacts to label noise differently. Results illustrated emphysema class was the most reactive to label noise compared to other classes. In contrast, the effusion class displayed the lowest change in performance due to added label noise. Taken together, multi-label classification models tolerated with up to 10% adding noise to multi classes and with up to 20% adding noise to single class additional error before performance was hindered. In the **binary disease/no disease** vs **multi-label classifier** comparison, the **binary disease/no disease** classifier outperformed the **multi-label** classification for focal disease classes (e.g., nodule). Importantly, performance as measured by AUC does not necessarily indicate that the **multi-label** disease classifier is worse. This is evidenced by the artifact/nature of our dataset where the nodule class overlap a lot with other diseases and **binary disease/no disease** model was able to pick those others easier, larger diseases but that was not actually nodule class performance.

In **binary disease/no disease** classification, nodule class overlap with other disease, and this **binary disease/no disease** model since this only looking at binary it is learning all disease at the same time, and no surprise that it is calling the case as positive when it happens to have a nodule, emphysema, and effusion. As results illustrated nodule class co-occurrence with emphysema always results in an improved AUC. And therefore, it even getting the credit for nodule, but it is really the emphysema that boosting the performance. Similarly, characteristics were of **binary nodule/no disease** classification is same as **binary disease/no disease**, whereas binary **nodule/ other classes** classification shown same level of performance as **multi-label** classification.

Another very interesting finding of our study as illustrated in fig. 8, **multi-label** model as effectively reacted to the co-occurrence of the clinically significant disease like atelectasis's association with effusion always reflect in a better identification of atelectasis in terms of AUC. Similar can be seen for emphysema and nodule class where nodule's association with emphysema always results in a better nodule performance.

A notable limitation of this study is that we haven't explored different approaches concerning explainable AI methodologies [31-34]. Our future efforts will investigate indemnifying features introducing bias in model decision making, clinically relevant relationships between co-occurring abnormalities. It will also be advantageous to understand the effect of co-occurrence disease morphologies and imaging properties in model decision-making. Mechanisms such as "concept whitening" can be applied to understand how the CNN model learned concept over the layers [32], systematic delectation of pathologies can also give a better understanding of the feature influences in decision making [34].

This paper analyzed a few interesting aspects of weakly supervised multi-label classification with an imbalanced and high-class co-occurrence dataset. Our finds related to label noise and model performance will encourage broadening the weakly supervised classification domain with noisy labels. Lastly, our observations on binary and multi-label classification demonstrate the pitfalls in training classifiers with co-occurring diseases and warn to be careful when interpreting the binary performance of particular classes where co-occurrence and prevalence of that class could be a highly impacting factor.

**Acknowledgment:**

We are grateful for the helpful discussions and data collection by Rui Hou and Brian Harrawood. This work was funded in part by developmental funds of the Duke Cancer Institute as part of the NIH/NCI P30 CA014236 Cancer Center Support Grant, the Center for Virtual Imaging Trials NIH/NIBIB P41-EB028744.